\documentclass{iopjournal}
\usepackage{amsmath}
\usepackage{amssymb}
\usepackage{dcolumn}
\usepackage{bm}
\hypersetup{colorlinks=true,citecolor=blue,linkcolor=black,urlcolor=blue}
\usepackage[super]{nth}
\usepackage{indentfirst}

\renewcommand{\articletype}[1]{}

\newcommand{\onlinecite}[1]{\cite{#1}}
\newcommand{\vek}[1]{\boldsymbol{#1}}
\newcommand{\uvek}[1]{\hat{\boldsymbol{#1}}}

\newcommand{\bnorm}{\uvek{b}}

\newcommand{\nperp}{\nabla_\perp}

\newcommand{\gav}[1]{\overline{#1}}
\newcommand{\Ah}{A_\parallel^{(h)}}
\newcommand{\As}{A_\parallel^{(s)}}

\newcommand{\bstar}{\boldsymbol{B}^\ast}

\newcommand{\pderiv}[1]{\frac{\partial}{\partial #1}}
\newcommand{\jacrz}{R \mathrm{d}R \mathrm{d}Z}
\newcommand{\jacrzr}{\mathrm{d}R \mathrm{d}Z}

\begin{document}
\articletype{Paper}
\title{A toroidally spectral field solver in the X-point Gyrokinetic Code for accurate simulation of reduced magneto-hydrodynamic modes}
\author{
Robert Hager$^{1,*}$\orcid{0000-0002-4624-3150},
C. S. Chang$^1$\orcid{0000-0002-3346-5731},
T. Gade$^2$\orcid{0009-0001-7401-5518},
E. Held$^3$\orcid{0000-0001-8695-7716},
S. Ku$^1$\orcid{0000-0002-9964-1208},
A. Mishchenko$^4$\orcid{0000-0003-1436-4502},
A. Scheinberg$^5$\orcid{0000-0002-5560-9716} and
B. Sturdevant$^6$\orcid{0000-0002-1116-2983}
}
\affil{$^1$Princeton Plasma Physics Laboratory, P.O. Box 451, Princeton, NJ 08543, USA}
\affil{$^2$University of Minnesota, 116 Church Street SE, Minneapolis, MN 55455, USA}
\affil{$^3$Fiat Lux, LLC, Logan, UT 84321, USA}
\affil{$^4$Max-Planck-Institut für Plasmaphysik, Wendelsteinstraße 1, 17491 Greifswald, Germany}
\affil{$^5$Jubilee Development, Cambridge, MA 02139, USA}
\affil{$^6$Rensselaer Polytechnic Institute, 110 8th Street, Troy, NY 12180, USA}
\affil{$^*$Author to whom any correspondence should be addressed.}
\email{rhager@pppl.gov}

{\fontsize{8}{10}\selectfont
\noindent This is the version of the article before peer review or editing, as submitted by the authors to \textit{Plasma Physics and Controlled Fusion}. IOP Publishing Ltd is not responsible for any errors or omissions in this version of the manuscript or in any version derived from it. If accepted for publication, the Version of Record will be available online in \textit{Plasma Physics and Controlled Fusion} via IOPscience.\par}


\begin{abstract}
A new field solver has been implemented in the global electromagnetic total-$f$ gyrokinetic particle-in-cell code XGC to extend the code's capability to large-scale reduced MHD-type instabilities in tokamak plasma.
While XGC's regular field solver is accurate at typical microturbulence scales of the order of the ion Larmor radius in tokamaks with arbitrary aspect ratio, a more accurate field solver is required for large-scale (i.e., low toroidal mode number) MHD-type modes such as internal kink, tearing and peeling modes.
The higher accuracy of the new field solver is achieved by dropping the (large aspect ratio) assumption that the poloidal magnetic field is much smaller than the toroidal magnetic field, while its numerical complexity is controlled by using a spectral discretization in the toroidal direction.
To cover the entire spectrum from large-scale MHD-type modes to small-scale microturbulence, the regular and the new field solver can be run alongside each other.
This work details the derivation of the new field solver, analyzes the differences between the XGC's regular and new field solvers, and verifies the new field solver against analytic predictions and the gyrokinetic code ORB5 and the MHD code NIMROD.
\end{abstract}


\section{Introduction}\label{sec:intro}
Multi-scale interaction of physics at different spatial scales is abundant in magnetically confined fusion plasma.
Of particular interest in this context are the interactions of large-scale magneto-hydrodynamic (MHD) instabilities with meso- and micro-scale gyrokinetic physics.

The objective of this work is to enable numerical studies of the interaction of MHD-type modes with neoclassical and micro-turbulence physics with a common set of equations using the global electromagnetic total-$f$ particle-in-cell X-point Gyrokinetic Code (XGC) \cite{xgc_code,ku_2018,Hager2022}.
While XGC has recently been extended to electromagnetic gyrokinetic physics \cite{Cole2021,Sturdevant2021,Hager2022}, in principle including reduced-MHD physics, XGC's solvers for the gyrokinetic Poisson equation and Amp\'{e}re's law (``field equations'') were, until now, optimized for gyrokinetic micro-turbulence and provided limited accuracy for large-scale MHD-type modes.
We present a new solver for XGC's field equations that achieves higher accuracy for large-scale modes.

Examples for the type of multi-scale physics targeted by this work are the interaction of gyrokinetic micro-turbulence in tokamaks with edge-localized modes (ELMs) or tearing modes, or the gyrokinetic plasma response to external three-dimensional (3D) magnetic perturbations (MPs).


ELMs \cite{leonard_2014} are periodic transport events caused by (MHD) peeling-ballooning instabilities in tokamak edge plasma.
They often eject a significant fraction of the plasma's stored energy, which can damage the plasma-facing components (PFCs) of the reactor, and must be avoided in a fusion pilot plant (FPP).
These ELM crashes have been observed to sometimes be preceded by metastable precursor modes \cite{diallo_2014,laggner_2016,burckhart_2010,lee_2016} with toroidal mode numbers in the peeling-ballooning range that coexist with micro-turbulence.
Recent research found evidence of three-wave interaction in DIII-D and KSTAR \cite{diallo_2018,Dominski2020,Kim2020}, and a theoretical model showed how a network of such nonlinear interactions might trigger an ELM crash \cite{Dominski2021}.

Unstable neoclassical tearing modes (NTMs) have been identified as the main cause of disruptions -- a catastrophic loss of plasma confinement -- in JET \cite{deVries2011}.
The large magnetic islands generated by the NTMs have been found to interact with micro-turbulence in experiments (e.g. Ref. \onlinecite{Choi2021}), and numerical simulations (e.g. Refs. \onlinecite{kwon_2018,Kwon2018_iaea,Fang2019,Orlov2023_iaea,Wei2024}).
Depending on whether turbulence can penetrate or is stabilized at the O-point of the magnetic island, the NTM can be stabilized or destabilized \cite{Choi2021}.

External MPs are a proven method for suppressing ELMs in tokamak plasma \cite{evans_2006,jeon_2012,suttrop_2018}, presumably by generating magnetic islands near the top of the pressure pedestal that limits the growth of the pedestal and prevents it from crossing the peeling-ballooning stability threshold \cite{nazikian_2015,hu_2019,hu_2020}.
The penetration of MPs into the plasma is routinely calculated with linear MHD and two-fluid approaches \cite{ferraro_2012,park_2007,liu_2011}, and sometimes with a quasilinear gyrokinetic model \cite{heyn_2008}.
While nonlinear MHD calculations of RMP penetration are being exercised, e.g., in Refs. \cite{hu_2019,hu_2020,Orain2016,Orain2019,Mitterauer2022,Sinha2022}, they remain computationally challenging, and kinetic effects like resistivity, viscosity, and heat conductivity are included with simple fluid closures, and turbulent transport is included via heuristic models.
Truly predictive understanding of ELM-suppression with MPs is still limited due to the complexity of the interplay between the MPs, the screening currents induced by the MPs in the plasma, micro-turbulence, and MHD-stability thresholds.

A common element of these examples is that (kinetic) neoclassical and turbulence effects in addition to MHD physics can influence the possibility to achieve a stable quiescent operating point in ITER and any FPP using the tokamak concept.
The convergence of MHD and kinetic physics described here is being approached from either side.
For example, MHD simulation codes use kinetic extensions to study energetic particle driven modes \cite{Liu2021, Liu2022,Liu2022_2,Todo1998,Todo2014}, or impurity transport with MPs \cite{Korving2024}.
And on the other hand, electromagnetic gyrokinetic codes are used to study microturbulence in the presence of MHD modes such as NTMs or MPs \cite{hager_2019_1,hager_2020,taimourzadeh_2019,holod_2016,kwon_2018,Wei2025,Yoo2022,Mishchenko2022,Navarro2017,Hornsby2010}, often with a fixed MHD perturbation.

Although reduced MHD and electromagnetic micro-turbulence physics have already been enabled in XGC \cite{Cole2021,Sturdevant2021,Hager2022}, the field solvers used in prior work were targeted at micro-turbulence, which has high toroidal mode numbers and varies slowly along the magnetic field (flute-like modes).
In this context, a large aspect ratio approximation was applied in the field equations that significantly reduced their numerical complexity by decomposing the originally 3D equations into a set of 2D equations.
We demonstrate that this approximation is adequate for the small spatial scales of micro-turbulence, especially in tokamak edge plasma, but can introduce significant errors for MHD modes with low toroidal mode numbers $n$.

To improve XGC's accuracy for low-$n$ electromagnetic modes, we introduce a new field solver that does not use the low-aspect ratio approximation.
By utilizing a toroidally spectral discretization, the new field solver remains 2D and can reuse the existing solver framework already implemented in XGC with few adjustments.
(Note that a low-$n$ solver for MHD-type modes has recently been developed also for the fluid-kinetic hybrid method in the gyrokinetic code GEM \cite{Zhang2022}.)
However, since combined simulation of MHD-type modes and micro-turbulence requires a broad range of toroidal mode numbers [$0<n<\mathcal{O}(100)$], the new field solver is applied in combination with the low aspect ratio solver to limit the total number of unknowns in the field equation.
We apply a Fourier filter to separate the low-$n$ ($0<n\lesssim 30)$ and high-$n$ ($n\gtrsim 30$) components of the spectrum, and solve the low-$n$ part of the spectrum with the new (spectral) solver, and the high-$n$ part with the conventional large aspect ratio solver.
A detailed error analysis comparing the new solver to the large aspect ratio solver justifies this approach.
Even for the low aspect ratio NSTX geometry, the difference between the two solvers becomes negligibly small for flute-like modes with $n\gtrsim 20$.
The difference is found to be especially large for modes with low poloidal mode numbers $m$, which, for flute-like modes, corresponds to low toroidal mode numbers $n$.

We show successful verifications of the new field solver for several key physics.
The frequency and damping rate of the shear Alfv\'{e}n wave is compared against an analytical formula and a different version of XGC that was introduced in Ref. \onlinecite{Sturdevant2021}.
The physics of the internal kink mode are benchmarked against the gyrokinetic code GTS \cite{Startsev2024}, and the MHD code NIMROD \cite{Held2015}, and the collisionless tearing mode is compared against the gyrokinetic code ORB5 \cite{Lanti2020}.
Finally, we also compare nonlinear electromagnetic turbulence in two XGC simulations using the new solver and the large aspect ratio solver, respectively.

The improved field equation solvers presented here enable accurate numerical studies with XGC of physics at the intersection of gyrokinetic and MHD physics.
While the interaction between turbulence and large magnetic islands or MPs in tokamak core plasma have been studied with various gyrokinetic codes before (see previous citations), this work extends these capabilities to tokamak edge plasma including the pedestal and scrape-off layer (SOL), and combining neoclassical, micro-turbulence and reduced MHD physics.
This is particularly relevant for ELM precursors and the onset of ELM crashes.

In the remainder of this paper, Sec. \ref{sec:equations} provides a brief introduction to XGC and discusses the large aspect ratio approximation used in XGC's field equations, including a qualitative error estimate.
Section \ref{sec:derivation} describes the toroidally spectral discretization of the field equations and their implementation in XGC.
A quantitative analysis of the differences between XGC's conventional large aspect ratio solver and the new solver developed in this work is presented in Sec. \ref{sec:error_analysis}.
The verification studies exercised for the new field solver are discussed in Sec. \ref{sec:verification}.
The paper closes with conclusions in Sec. \ref{sec:conclusions}.

\section{The Turbulence Approximation in the Gyrokinetic Field Equations}\label{sec:equations}

\subsection{The X-point Gyrokinetic Code (XGC)}\label{subsec:xgc_intro}
The X-point Gyrokinetic Code (XGC) \cite{xgc_code} is a global electromagnetic gyrokinetic particle-in-cell code for the simulation of magnetically confined fusion plasma.
XGC is specialized for simulation of plasma in the pedestal, edge and scrape-off layer of magnetic fusion devices, where the strong pressure gradients and the vicinity of a material wall can make edge plasma deviate substantially from a Maxwellian distribution.

XGC's defining features are the use of a total-$f$ method \cite{sku_2016} that is required for simulation of highly non-Maxwellian plasma;
the capability to simulate the whole volume of tokamaks enabled by the use of approximately field-aligned unstructured meshes \cite{adams_ku_2009,fzhang_2015,Riaz2024} in cylindrical coordinates; a logical sheath wall boundary condition;
an implicit, nonlinear Fokker-Planck collision operator \cite{hager2016_2,yoon_2014};
Monte-Carlo neutral recycling \cite{stotler_2017};
support for multiple impurity particle species \cite{Dominski2019,Dominski2024};
and source routines for external heat and torque input.

Detailed descriptions of the capabilities and models used in XGC are given in Refs. \onlinecite{ku_2018,Hager2022}.
To provide context for the remainder of the article, we only reference the equations of motion of the marker particles, and the corresponding field equations.
XGC's governing equation is the five-dimensional (5D) gyrokinetic Boltzmann equation
\begin{equation}\label{eq:xgc_base_eq}
    \frac{\partial f}{\partial t} + \dot{\vek{z}} \cdot \frac{\partial f}{\partial \vek{z}} = S(f), 
\end{equation}
where $\vek{z}=(\vek{R},u_\parallel,\mu_s)$, $\vek{R}=(R,\varphi,Z)$, and $\mu_s$ is the magnetic moment of a particle of species $s$.
The left-hand-side is solved with a particle-in-cell method.
The right-hand-side represents all sources and sinks such as collisions, neutral particle recycling, etc., and is solved with continuum methods on a 5D phase space mesh.

The Lagrangian equations of motion ($\dot{\vek{z}}$) are \cite{Kleiber_2016}
\begin{align}
\dot{\vek{R}} &= \frac{D}{B_0} \left[ \frac{\bstar}{m_s}\,\frac{\partial H}{\partial u_\parallel} + \frac{\vek{F}\times\vek{B}_0}{B_0} \right], \label{eq:xgc_rdot} \\
\dot{u}_\parallel &= \frac{q_s}{m_s} \left[ D \frac{\bstar}{B_0} \cdot \vek{F} - \frac{\partial \As}{\partial t} \right], \label{eq:xgc_udot}
\end{align}
where $\vek{B}_0=\nabla \times \vek{A}_0$ is the equilibrium magnetic field, $B_0=|\vek{B}_0|$, $\bnorm = \vek{B_0}/B_0$, $\bstar=\nabla \times \vek{A}^\ast$ and
\begin{align}
\vek{A}^\ast &= \vek{A}_0 + \left(\frac{m_s}{q_s} u_\parallel + \gav{\As}\right) \bnorm , \label{eq:xgc_astar}\\
H &= \frac{m_s}{2} u_\parallel^2 + \mu_s B_0 + q_s \left( \gav{\phi} - u_\parallel \gav{\Ah} \right) + \frac{q_s^2}{2 m_s} \gav{\Ah}^2, \label{eq:xgc_hamiltonian} \\
\frac{\partial H}{\partial u_\parallel} &= m u_\parallel - q_s \Ah, \\
\vek{F} &= -\frac{1}{q_s}\,\frac{\partial H}{\partial \vek{R}} = -\frac{\mu_s}{q_s} \nabla B_0 - \nabla\gav{\phi} + u_\parallel \nabla\gav{\Ah} - \frac{q_s}{m_s} \gav{\Ah} \left( \nabla \gav{\Ah} \right), \label{eq:xgc_ham_force} \\
D &= \frac{B_0}{\bnorm \cdot \bstar} = \left[ 1 + \left( \frac{m_s u_\parallel}{q_s B_0} + \frac{\gav{\As}}{B_0} \right) \bnorm \cdot \nabla \times \bnorm \right]^{-1}. \label{eq:xgc_D}
\end{align}
In equations \eqref{eq:xgc_rdot} to \eqref{eq:xgc_D}, $m_s$ is the mass and $q_s$ the charge of species $s$, and $u_\parallel=v_\parallel+ (q_s/m_s) \gav{\Ah}$ is the ``canonical'' parallel velocity, $\mu_s = m_s v_\perp^2/(2 B_0)$ is the magnetic moment.
The complete physical vector potential is given by $A_\parallel = \vek{A}\cdot\bnorm = \As+\Ah$.
An overline $\overline{A} = 1/(2\pi) \oint A(\boldsymbol{R}+\boldsymbol{\rho}) \mathrm{d} \alpha$ indicates the gyroaverage.

In the long wavelength limit, the system is closed by the gyrokinetic Poisson equation and Amp\'{e}re's law:

\begin{align}
 \nabla\cdot \frac{n_0 m_i}{B_0^2} \nabla_\perp \phi &= -\left(q_i \overline{n}_i + q_e n_e \right) \label{eq:xgc_gk_poisson} \\
 -\nabla \cdot \nabla_\perp \Ah + \Ah \sum_{s=i,e} \frac{\mu_0 n_0 q_s^2}{m_s}
      &= \mu_0 \left( \gav{\delta j_{\parallel,i}} + \delta j_{\parallel,e} \right) + \nabla \cdot \nabla_\perp \As, \label{eq:xgc_ampere}
\end{align}
where $n_0$ and $T_{i/e}$ are the background density and ion/electron temperatures, $\rho_i=(m_i k_B T_i)^{1/2} /(q_i B_0)$ is the ion gyroradius, $\mu_0$ is the permeability of free space, and the parallel current densities $j_{\parallel,s}$ are given by the first $u_\parallel$ moments of the species distribution functions $f_s$.

The discretization of phase space $(R,\varphi,Z,u_\parallel,\mu)$ used by XGC is explained in more detail in Ref. \onlinecite{Hager2022}, and the configuration space mesh generation in Ref. \onlinecite{Riaz2024}.
Here, we briefly review only the discretization of configuration space $(R,\varphi,Z)$ because it is essential for understanding the derivation and implementation of the solver presented in this work.
XGC uses a uniform discretization in the toroidal angle \cite{Hager2022} with the computational domain being $0\leq \varphi \leq 2\pi/N_w$, where $N_w$ is the (integer) wedge number, grid size $N_\varphi$, and resolution $\Delta\varphi = 2\pi/(N_w N_\varphi)$.
Periodic boundary conditions are applied in the toroidal direction, if $N_w>1$.
Each poloidal plane $(R,\varphi_i=i \Delta \varphi,Z)$, is discretized with an unstructured triangle mesh with its vertices approximately aligned with the magnetic field $\vek{B}_0$.
That is, a magnetic field-line starting at a mesh vertex on plane $\varphi_i$ intersects plane $\varphi_{i+1}$ at or very close to another mesh vertex.
Any interpolation between adjacent poloidal planes (e.g., scatter-gather operations, gradient calculation) is executed along the magnetic field instead of the toroidal direction.
Therefore, XGC can resolve perturbations with high toroidal mode number ($n_{max} > N_w N_\varphi/2$), with relatively low toroidal resolution $N_\varphi$, if the perturbations vary slowly along the magnetic field ($k_\parallel \ll k_\perp$), as is the case for magnetic confinement fusion plasma.
However, a sufficiently high resolution on the radial-poloidal planes is required to resolve the small scales of microturbulence perpendicular to the magnetic field.
In this regard, XGC's discretization is similar to the flux-coordinate independent approach (FCI) \cite{Hariri2013} employed in magnetic confinement plasma codes using continuum methods, e.g., GENE-X \cite{Michels2021}, or BSTING \cite{Shanahan2018}.

\subsection{Approximations in the Field Equations}\label{subsec:field_eq_approx}
Before this work, XGC has been focused on micro-turbulence in tokamak edge plasma, where the magnetic safety factor $q$, a measure for the average field-line pitch (ratio of toroidal to poloidal magnetic field), and toroidal mode numbers $n$ are both high.
In this context, a convenient approximation can be used to simplify solving the field equations \eqref{eq:xgc_gk_poisson} and \eqref{eq:xgc_ampere}.
We discuss this approximation in this section.

We use a right-handed cylindrical coordinate system $\xi^i = (R,\varphi,Z)$ and basis vectors $\nabla \xi^i$.
For simplifying equations and unless noted otherwise, we use a sum convention so that $\sum_i a_i a^i$ is contracted to $a_i a^i$.
The metric tensor is given by
\begin{equation}
    \left( g_{ij} \right) =
    \begin{pmatrix}
        1 & 0   & 0 \\
        0 & R^2 & 0 \\
        0 & 0   & 1
    \end{pmatrix}
\end{equation}
with the Jacobian $1/\mathcal{J}=\sqrt{\det(g_{ij})} = R$.

The field equations \eqref{eq:xgc_gk_poisson} and \eqref{eq:xgc_ampere} can be written in the generalized form
\begin{equation}\label{eq:exact_pol}
    \left[ -\nabla \cdot \alpha(R,Z) \nabla_\perp + n\beta(R,Z) \right] \phi(R,\varphi,Z) = \rho(R,\varphi,Z),
\end{equation}
where the perpendicular gradient $\nabla_\perp$ is given by
\begin{equation}\label{eq:perp_grad}
    \nabla_\perp=\nabla-\bnorm \left(\bnorm \cdot \nabla \right) = \nabla \xi^i \left( \frac{\partial}{\partial \xi^i} - \frac{B_i B^j}{B^2} \,\frac{\partial}{\partial \xi^j} \right).
\end{equation}

The presence of all three derivatives with respect to $R$, $\varphi$, and $Z$, implies that any discretization of the field equations would normally have to be 3D.
However, the numerical complexity of solving the field equations can be reduced by splitting the original 3D problem into a set of 2D problems.

This can be achieved without approximation by using a discretization on curved planes that are perpendicular to the magnetic field.
For example, in gyrokinetic codes using flux-coordinates $(x,y,z)$, $B^x=0=B^y$ (e.g., Ref. \onlinecite{Scott2001}), and $B^2 = B_z B^z$ so that all the derivatives with respect to $z$ drop out of $\nabla \cdot \alpha \nabla_\perp$:
\begin{equation}\label{eq:polarization_flux_coords}
    \nabla \cdot \alpha \nabla_\perp = \mathcal{J} \left\lbrace
        \frac{\partial}{\partial x} \left[ \mathcal{J}^{-1} \left( g^{xx} \frac{\partial}{\partial x} + g^{xy} \frac{\partial}{\partial y} \right) \right]
        +\frac{\partial}{\partial y} \left[ \mathcal{J}^{-1} \left( g^{yx} \frac{\partial}{\partial x} + g^{yy} \frac{\partial}{\partial y} \right) \right]
    \right\rbrace,
\end{equation}
where $\mathcal{J}$ is the Jacobian.

Another commonly used option, e.g. in GENE-X \cite{Michels2021}, and, before this work, in XGC, is to apply the large aspect ratio approximation, or $\epsilon \ll 1$ with the inverse aspect ratio $\epsilon$.
In this limit, the poloidal magnetic field is much smaller than the toroidal one, which implies that
\begin{equation}\label{eq:b_ordering}
    \frac{|B_R|}{B} \sim \frac{|B_Z|}{B} \sim \epsilon \ll 1 \Leftrightarrow |B_\phi| \sim B.
\end{equation}
By dropping all terms of order $\epsilon$ or higher in Eq. \eqref{eq:perp_grad}, the perpendicular gradient $\nabla_\perp$ simplifies to
\begin{equation}
    \nabla_\perp \approx \nabla R \frac{\partial}{\partial R} + \nabla Z \frac{\partial}{\partial Z},
\end{equation}
so that the first term on the left hand side of Eq. \eqref{eq:xgc_gk_poisson} becomes
\begin{equation}\label{eq:approx_pol}
    -\nabla \cdot \alpha \nabla_\perp \phi \approx -\left[ \frac{1}{R}\,\frac{\partial}{\partial R} \left( R \alpha \frac{\partial}{\partial R} \right) + \frac{\partial}{\partial Z} \left( \alpha \frac{\partial}{\partial Z} \right) \right] \phi.
\end{equation}
%

Since Eq. \eqref{eq:approx_pol} contains only derivatives with respect to $R$ and $Z$, the original 3D field equations \eqref{eq:xgc_gk_poisson} and \eqref{eq:xgc_ampere} decompose into a set of 2D equations when discretized in the toroidal direction, a significant simplification.

\subsection{Error estimate}\label{subsec:polarization_error_estimate}

It is worthwhile to estimate the error introduced through this approximation to understand the limits of its applicability in situations with finite poloidal magnetic field.
In terms of the radial-poloidal wavenumber $\vek{k}_{R,Z}$, the toroidal wavenumber $k_\varphi$, and the parallel wavenumber $k_\parallel$, we can formally express the perpendicular gradient, Eq. \eqref{eq:perp_grad}, of a test function $\phi$ as
\begin{align}\label{eq:k_perp_estimate1}
    \frac{\nperp \phi}{\phi} &= \vek{k}_\perp \approx \boldsymbol{k}_{RZ} + k_\varphi \uvek{\varphi} - k_\parallel \bnorm \notag \\
    &= \left(\vek{k}_r + \vek{k}_\theta \right) + k_\varphi \uvek{\varphi} - k_\parallel \bnorm.
\end{align}
Assuming $k_r \ll 1$, and using the following order-of-magnitude estimates,
\begin{align}
    k_\parallel &\sim \frac{n_\parallel}{R_0} = \frac{m/q - n}{R_0}, \label{eq:k_para_estimate}\\
    k_\theta &\sim \frac{m}{r} \sim \frac{(n-n_\parallel) q}{\epsilon R_0}, \label{eq:k_theta_estimate}\\
    k_\varphi &\sim \frac{n}{R_0}, \label{eq:k_tor_estimate}
\end{align}
where $n$ and $m$ are the toroidal and poloidal mode numbers, $R_0$ is the major radius, and $q$ is the safety factor, we obtain the following expression for the order of magnitude of the perpendicular wavenumber $k_\perp$
\begin{equation}\label{eq:k_perp_estimate_full}
    k_\perp^2 \sim \left( \frac{(n-n_\parallel) q}{\epsilon R_0} \right)^2 + \frac{n^2}{R_0^2} + \frac{n_\parallel^2}{R_0^2} - 2 \frac{n_\parallel}{R_0^2} \left( (n-n_\parallel) \frac{q B_{pol}}{\epsilon B} + n \frac{B_{tor}}{B} \right).
\end{equation}
For the approximated perpendicular wavenumber according to Eq. \ref{eq:approx_pol}, we find
\begin{equation}\label{eq:k_perp_estimate_approx}
    \tilde{k}_\perp^2 \sim \left( \frac{(n-n_\parallel) q}{\epsilon R_0} \right)^2.
\end{equation}

For exactly field-aligned modes ($m=nq$ or $n_\parallel=0$), and separately in the limit of large toroidal mode numbers ($n \gg n_\parallel$), the relative difference between the approximated and exact $k_\perp$ becomes
\begin{equation}\label{eq:k_perp_error}
    \Delta k_\perp = \sqrt{\frac{\left( \tilde{k}_\perp - k_\perp \right)^2}{k_\perp^2}} \sim \left| 1-\frac{|q|}{\sqrt{\epsilon^2+q^2}} \right|.
\end{equation}
%
This error becomes small not only at small inverse aspect ratio, but also at finite inverse aspect ratio and sufficiently large safety factor, as is typical for tokamak edge plasma.
Our analysis shows that the large aspect ratio approximation can be acceptable for micro-turbulence in tokamak edge plasma in many situations.

However, for modes with $m=0$ (and $k_r\ll1$), $\tilde{k}_\perp$ approaches zero, and the corresponding relative difference between the approximate and accurate $k_\perp$ is of order unity.
Consequently, we expect the largest errors for modes with finite $k_\parallel$ and low toroidal mode number $n$, i.e., exactly the modes that are relevant for MHD-type instabilities such as peeling-ballooning modes.

We caution the reader, however, that the result in Eq. \eqref{eq:k_perp_error} is not a fully quantitative error analysis, but rather an estimate that serves to illustrate the asymptotic behavior of the approximate field equation.
A quantitative error analysis is presented in Sec. \ref{sec:error_analysis}.

\section{Derivation of a toroidally spectral field solver}\label{sec:derivation}
From the error estimate in Sec. \ref{subsec:polarization_error_estimate}, it is obvious, that XGC's field solvers, which use Eq. \eqref{eq:approx_pol}, must be improved.
The least invasive option for implementing the full perpendicular gradient operator in Eq. \eqref{eq:exact_pol} in XGC is to use a spectral discretization in the toroidal direction together with a finite-element discretization on the radial-poloidal planes.
This approach preserves the decomposition of the 3D field equations \eqref{eq:xgc_gk_poisson} and \eqref{eq:xgc_ampere} into a set of 2D equations, and it allows us to continue using the same solver methods already established in XGC.
(We will explain those methods in Secs. \ref{subsec:fem_discretization} and \ref{subsec:petsc_implementation}.)

In the following, we describe the discretization of the gyrokinetic field equations using a toroidally spectral discretization, and the methods to invert the discretized equations.

\subsection{Toroidally spectral discretization}\label{subsec:spectral_discretization}

The first step in the discretization of Eq. \eqref{eq:exact_pol} is a toroidal Fourier decomposition.
Using $\nabla \cdot \vek{A} = \mathcal{J} \pderiv{\xi^i} (A^i/\mathcal{J})$ ,and an arbitrary test function $\phi$, the explicit form of the general field equation is
\begin{align}
    -\nabla \cdot \left( \alpha \nabla_\perp \phi \right) + \beta\phi =
       \frac{1}{R} \,\pderiv{R} &\left[ \alpha R \left( \pderiv{R}\phi -\frac{B_R B^j}{B^2} \pderiv{\xi^j} \phi \right) \right] \notag\\
    + \pderiv{Z} &\left[ \alpha \left( \pderiv{Z}\phi -\frac{B_Z B^j}{B^2} \pderiv{\xi^j} \phi \right) \right] \notag\\
    + \frac{1}{R^2} \pderiv{\varphi} &\left[ \alpha \left( \pderiv{\varphi}\phi -\frac{B_\varphi B^j}{B^2} \pderiv{\xi^j} \phi \right) \right] + \beta\phi = \rho. \label{eq:field_equation_explicit}
\end{align}
We will refer to the first term on the left hand side of Eq. \eqref{eq:field_equation_explicit} as the ``polarization term'' because it represents the polarization density in the gyrokinetic Poisson equation.
The terms $\beta\phi$ and $\rho$ will be called ``mass terms'' (following standard finite-element terminology) because they contain no differential operators.

In the following derivation, we assume $N_w=1$ for simplicity; the generalization to $N_w>1$ is straightforward.
The functional space of periodic test functions on a uniform toroidal angle grid can be represented as a Fourier series,
\begin{equation}\label{eq:phi_fourier_def}
    \phi = \sum_{n=0}^{n_{max}} \left( \phi_n^{(s)} \sin(n\varphi) + \phi_n^{(c)} \cos(n\varphi) \right) = \sum_{n=0}^{n_{max}} \phi_n.
\end{equation}
%
Note that in most cases $n_{max} > N_\varphi/2$ as explained in Sec. \ref{subsec:high-n_solver}.

After substituting Eq. \eqref{eq:phi_fourier_def} into Eq. \eqref{eq:field_equation_explicit}, we can multiply the field equation by $\sin(n\varphi)$ and $\cos(n\varphi)$, integrate over the toroidal angle, and explicitly evaluate all derivatives with respect to $\varphi$ to obtain a system of two equations for the sine and cosine components of the field equation:
\begin{align}
    -\left[ \nabla \cdot \left( \alpha \nabla_\perp \phi_n^{(s)} \right) \right] &+ \beta \phi_n^{(s)} = \notag\\
    &-\frac{1}{R} \pderiv{R} \left\lbrace \alpha R \left[ \left( 1-\frac{B_R^2}{B^2} \right) \pderiv{R} \phi_n^{(s)} -\frac{B_R B_Z}{B^2} \pderiv{Z} \phi_n^{(s)} + n \frac{B_R B^\varphi}{B^2} \phi_n^{(c)} \right]  \right\rbrace \notag\\
    &-\pderiv{Z} \left\lbrace \alpha \left[ \left( 1-\frac{B_Z^2}{B^2} \right) \pderiv{Z} \phi_n^{(s)} - \frac{B_R B_Z}{B^2} \pderiv{R} \phi_n^{(s)} + n \frac{B_Z B^\varphi}{B^2} \phi_n^{(c)} \right] \right\rbrace \notag\\
    &+ \left\lbrace \frac{\alpha}{R^2} \left[ n^2 \left( 1-\frac{B_\varphi B^\varphi}{B^2} \right) \phi_n^{(s)} - n \frac{B_R B_\varphi}{B^2} \pderiv{R} \phi_n^{(c)} - n \frac{B_Z B_\varphi}{B^2} \pderiv{Z} \phi_n^{(c)} \right] \right\rbrace \notag\\
    &= \rho_n^{(s)},\label{eq:poisson_spec_sin}\\
    -\left[ \nabla \cdot \left( \alpha \nabla_\perp \phi_n^{(c)} \right) \right] &+ \beta \phi_n^{(c)} = \notag\\
    &-\frac{1}{R} \pderiv{R} \left\lbrace \alpha R \left[ \left( 1-\frac{B_R^2}{B^2} \right) \pderiv{R} \phi_n^{(c)} -\frac{B_R B_Z}{B^2} \pderiv{Z} \phi_n^{(c)} - n \frac{B_R B^\varphi}{B^2} \phi_n^{(s)} \right]  \right\rbrace \notag\\
    &-\pderiv{Z} \left\lbrace \alpha \left[ \left( 1-\frac{B_Z^2}{B^2} \right) \pderiv{Z} \phi_n^{(c)} - \frac{B_R B_Z}{B^2} \pderiv{R} \phi_n^{(c)} - n \frac{B_Z B^\varphi}{B^2} \phi_n^{(s)} \right] \right\rbrace \notag\\
    &+ \left\lbrace \frac{\alpha}{R^2} \left[ n^2 \left( 1-\frac{B_\varphi B^\varphi}{B^2} \right) \phi_n^{(c)} + n \frac{B_R B_\varphi}{B^2} \pderiv{R} \phi_n^{(s)} + n \frac{B_Z B_\varphi}{B^2} \pderiv{Z} \phi_n^{(s)} \right] \right\rbrace \notag\\
    &= \rho_n^{(c)}.\label{eq:poisson_spec_cos}
\end{align}
%
%

Like the large large aspect ratio approximation discussed in Sec. \ref{subsec:field_eq_approx}, the toroidally spectral discretization applied here decomposes the 3D field equation into a set of 2D equations.
However, the number of unknowns per 2D problem is twice as large as in case of Eq. \eqref{eq:approx_pol}.

\subsection{Linear finite-element discretization on the poloidal plane}\label{subsec:fem_discretization}
On the radial-poloidal planes $\varphi=\mathrm{const.}$ (or for each toroidal Fourier mode), we employ a linear finite-element discretization with the basis functions defined as
%
\begin{equation}\label{eq:linear_elements}
    \Lambda_i = \sum_j \lambda_{ij} = \sum_j \left( a_{ij} R + b_{ij} Z + c_{ij} \right),
\end{equation}
where the index $i$ is the mesh-vertex index and the sum is over all elements (triangles) $j$ that have the vertex $i$ in common.
The linear basis function have compact support, i.e., $\lambda_{ij}=0$ if $(R,Z) \notin \Omega_j$, where $\Omega_j$ is the surface of triangle $j$. 
The volume element for the integration over $R$ and $Z$ is $R \mathrm{d}R \mathrm{d}Z$.

To derive the weak form of field equations, we multiply Eqs. \eqref{eq:poisson_spec_sin} and \eqref{eq:poisson_spec_cos} by $\Lambda_l$, substitute $\phi_n^{(c/s)}=\sum_i \phi_{ni}^{(c/s)} \Lambda_i$, and integrate over $R$ and $Z$.
Using integration by parts, and $\lambda_{ij} \lambda_{lm}=0$ for $j \neq m$ (different elements), we obtain for the polarization term

\begin{align}
    \int \Lambda_l \left[-\nabla \cdot \left( \alpha \nabla_\perp \phi_n^{(s)} \right) \right] & \jacrz \notag\\
    = \sum_{i,j} \int &\left\lbrace \left[ \alpha R \left( \left( 1-\frac{B_R^2}{B^2} \right) a_{ij} a_{lj} + \left( 1-\frac{B_Z^2}{B^2} \right) b_{ij} b_{lj} \right. \right. \right. \notag\\
    &+\left. \left. \left. \frac{n^2}{R^2} \left( 1-\frac{B_\varphi B^\varphi}{B^2} \right) \lambda_{ij} \lambda_{lj} - \frac{B_R B_Z}{B^2} \left( a_{ij} b_{lj} + b_{ij} a_{lj} \right) \right) \right] \phi_{ni}^{(s)} \right. \notag\\
    &+ \left[ \alpha R n \left( \frac{B_R B^\varphi}{B^2} \left( \lambda_{ij} a_{lj} - a_{ij} \lambda_{lj} \right) \right. \right.  \notag\\
    &+\left. \left. \left. \frac{B_Z B^\varphi}{B^2} \left( \lambda_{ij} b_{lj} - b_{ij} \lambda_{lj} \right) \right) \right] \phi_{ni}^{(c)} \right\rbrace \jacrzr,\label{eq:poisson_spec_sin3}\\
    \int \Lambda_l \left[-\nabla \cdot \left( \alpha \nabla_\perp \phi_n^{(c)} \right) \right] & \jacrz \notag\\
    = \sum_{i,j} \int &\left\lbrace \left[ \alpha R \left( \left( 1-\frac{B_R^2}{B^2} \right) a_{ij} a_{lj} + \left( 1-\frac{B_Z^2}{B^2} \right) b_{ij} b_{lj} \right. \right. \right. \notag\\
    &+\left. \left. \left. \frac{n^2}{R^2} \left( 1-\frac{B_\varphi B^\varphi}{B^2} \right) \lambda_{ij} \lambda_{lj} - \frac{B_R B_Z}{B^2} \left( a_{ij} b_{lj} + b_{ij} a_{lj} \right) \right) \right] \phi_{ni}^{(c)} \right. \notag\\
    &- \left[ \alpha R n \left( \frac{B_R B^\varphi}{B^2} \left( \lambda_{ij} a_{lj} - a_{ij} \lambda_{lj} \right) \right. \right.  \notag\\
    &+\left. \left. \left. \frac{B_Z B^\varphi}{B^2} \left( \lambda_{ij} b_{lj} - b_{ij} \lambda_{lj} \right) \right) \right] \phi_{ni}^{(s)} \right\rbrace \jacrzr.\label{eq:poisson_spec_cos3}
\end{align}
And the weak form of the mass terms is
\begin{equation}\label{eq:mass_terms_fem}
    \int \beta \Lambda_l \rho_n^{(c/s)} \jacrz = \sum_{i,j} \int \beta \rho_{ni}^{(c/s)} \lambda_{ij} \lambda_{lj} \jacrz.
\end{equation}
For the regular mass matrix, $\beta=1$, and we will use the symbol $M$.
For the simplest form of the adiabatic response term in the Poisson equation \eqref{eq:xgc_gk_poisson}, $\beta=n_0/T_e$, and for the skin-term in Amp\'{e}re's law \eqref{eq:xgc_ampere}, $\beta=\sum_{s=i,e} \frac{\mu_0 n_0 q_s^2}{m_s}$.
We will refer to these terms with the common symbol $M_{ad}$.

For a compact representation of the resulting field equation, we define
\begin{align}
    A_n &= \sum_{j} \int \left\lbrace \alpha R \left[ \left( 1-\frac{B_R^2}{B^2} \right) a_{ij} a_{lj} + \left( 1-\frac{B_Z^2}{B^2} \right) b_{ij} b_{lj} \right. \right. \notag\\
    &+\left. \left. \frac{n^2}{R^2} \left( 1-\frac{B_\varphi B^\varphi}{B^2} \right) \lambda_{ij} \lambda_{lj} - \frac{B_R B_Z}{B^2} \left( a_{ij} b_{lj} + b_{ij} a_{lj} \right) \right] \right\rbrace \jacrzr,\label{eq:pol_op_spectral_a} \\
    B_n &= \sum_j \int \left\lbrace \alpha R n \left[ \frac{B_R B^\varphi}{B^2} \left( \lambda_{ij} a_{lj} - a_{ij} \lambda_{lj} \right) + \frac{B_Z B^\varphi}{B^2} \left( \lambda_{ij} b_{lj} - b_{ij} \lambda_{lj} \right) \right] \right\rbrace \jacrzr, \label{eq:pol_op_spectral_b} \\
    M_\beta &= \sum_j \int \beta R \lambda_{ij} \lambda_{lj} \jacrzr.\label{eq:mass_matrix}
\end{align}
With these definitions, the discretized field equation is
\begin{equation}
    \left[
    \begin{pmatrix}
       A_n & B_n \\
      -B_n & A_n
    \end{pmatrix}
    +
    \begin{pmatrix}
       M_{ad} & 0 \\
       0 & M_{ad}
    \end{pmatrix}
    \right]
    \cdot
    \begin{pmatrix}
      \phi_n^{(s)} \\
      \phi_n^{(c)}
    \end{pmatrix}
    =
    \begin{pmatrix}
       M & 0 \\
       0 & M
    \end{pmatrix}
    \cdot
    \begin{pmatrix}
      \rho_n^{(s)} \\
      \rho_n^{(c)}
    \end{pmatrix}
    . \label{eq:block_matrix_system}
\end{equation}

This result reduces to the approximated field equations as previously implemented in XGC when applying the ordering in Eq. \eqref{eq:b_ordering} and retaining only the lowest order terms:
\begin{align}
    A &= \sum_{j} \int \left[ \alpha R \left(  a_{ij} a_{lj} +  b_{ij} b_{lj} \right) \right] \jacrzr \label{eq:a_op_approx} \\
    B &= 0.\label{eq:b_op_approx}
\end{align}
Since the sine and cosine components are not coupled in this approximation, and $A$ is independent of $n$, all toroidal mode numbers can be solved together from a single equation.

%


\subsection{Solving for high toroidal mode numbers}\label{subsec:high-n_solver}
%

A toroidally spectral discretization of the gyrokinetic field equations usually requires solving one equation for each toroidal mode number retained in the simulation.
In a conventional aspect ratio tokamak with a size comparable to DIII-D \cite{diii-d_design}, a minor radius $a \sim 0.6$ m, an edge gyroradius of $\rho_i \sim 2$ mm, an edge safety factor $q \sim 4$, and turbulence on the gyroradius scale, $k_\perp \rho_i \sim 1$ ($\lambda \sim 2\pi \rho_i$), the upper limit of the toroidal mode number is at least $n \sim a/(q \rho_i) = 75$.
This is relatively high compared to typical MHD simulations and often larger than the toroidal resolution $N_\varphi$.

To limit the number of equations to be solved while maintaining adequate accuracy, we use the accurate solver, Eq. \eqref{eq:block_matrix_system} with Eqs. \eqref{eq:pol_op_spectral_a} and \eqref{eq:pol_op_spectral_b}, only for low toroidal mode numbers $0< n \leq N_\varphi$
that are relevant for MHD-type modes.
For higher mode numbers $n > N_\varphi$, we use the field solver in the large aspect ratio approximation, Eq. \eqref{eq:block_matrix_system} with Eqs. \eqref{eq:a_op_approx} and \eqref{eq:b_op_approx}.
The resulting number of field equations to be solved in this approach is $2 N_\varphi$, compared to $N_\varphi$ field equations when using the large aspect ratio solver alone.
The total number of unknowns is $3 N_\varphi N_n$ compared to $N_\varphi N_n$, where $N_n$ is the number of mesh vertices per poloidal plane.

To separate the various toroidal mode numbers a spectral discretization requires field-aligned (2D) Fourier filtering in the straight-field line poloidal angle $\theta^\ast$ [Eq. (22) in Ref. \onlinecite{Hager2022}] and the toroidal angle $\varphi$ of the right-hand-side and the solution of the field equation.
The reason why a 2D Fourier filter is required instead of a simple 1D toroidal one is discussed in Sec. \ref{subsec:xgc_intro} and illustrated in Ref. \onlinecite{Hariri2015}.
While the field-aligned construction of XGC's configuration space mesh and the field-aligned evaluation of $\bnorm \cdot \nabla$, allow for accurate representation of toroidal mode numbers $n>N_\varphi$, those high toroidal mode numbers appear as faux-aliases of modes with $n_{num} \leq N_\varphi$ in a poloidal-toroidal mode spectrum.

The solver and filter algorithm starts with the spectral decomposition of the right-hand-side vector $\rho$.
The first step is a 1D toroidal Fourier transform, which yields the sine and cosine Fourier coefficients $\rho_{n,num}^{(s)}(R,Z)$ and$\rho_{n,num}^{(c)}(R,Z)$, where
\begin{equation}\label{eq:real-to-num_mode}
    n_{num} = \left\lbrace\begin{matrix}
                 n - \frac{i N_\varphi}{2} & \text{even } i\\
                 \frac{N_\varphi}{2} - \left(n - \frac{i N_\varphi}{2} \right) & \text{odd } i
               \end{matrix}\right.,
\end{equation}
with $i=\lfloor 2(n-1)/N_\varphi \rfloor$ takes into account the faux-aliasing mentioned above.
Next, a field-aligned poloidal Fourier filter that retains only the contributions of poloidal mode numbers that satisfy $|m/q-n|<N_\parallel^{max}<N_\varphi/2$ is applied to the Fourier coefficients $\rho_{n,num}^{(s)}$ and $\rho_{n,num}^{(c)}$ to obtain $\rho_{n}^{(s)}$ and $\rho_{n}^{(c)}$.
Note that this filter, in order to be able to isolate field-aligned modes, is made sensitive to the sign of the poloidal mode number $m$ using symmetry properties of the Fourier transform.
The right-hand-side vector $\rho_{n>N_\varphi}$ for the high toroidal mode number field equation is obtained by applying a 2D field-aligned Fourier filter to the full right-hand-side $\rho$ to remove all contributions from modes $n\leq N_\varphi$.

After solving the field equations, the same field-aligned poloidal Fourier filter is applied to the solution $\phi_{n}^{(s)}$ and $\phi_{n}^{(c)}$ of the low toroidal mode number field equation, and the 2D field-aligned Fourier filter is applied to the solution $\phi_{n>N_\varphi}$ of the high toroidal mode number equation.
The full solution of the field equation is then given by the sum of the low and high toroidal mode number solutions
\begin{equation}\label{eq:combined_field_solution}
    \phi = \sum_{n=1}^{N_\varphi} \left[\phi_{n}^{(c)} \cos(n\varphi) + \phi_{n}^{(s)} \sin(n\varphi) \right] + \phi_{n>N_\varphi}.
\end{equation}


\subsection{Implementation with PETSc iterative solvers and parallel performance}\label{subsec:petsc_implementation}
%

To solve the system in Eq. \eqref{eq:block_matrix_system}, we use methods available from the PETSc library \cite{petsc-web-page,petsc-user-ref,petsc-efficient}.
We treat the block matrix on the left-hand side monolithically, employing a preconditioned conjugate gradient method for the outer iterations.
Note that the block matrix system is symmetric positive definite, which is a prerequisite for using conjugate gradient methods.
For preconditioning, we use a single application of a V(2,2) algebraic multigrid (AMG) cycle with symmetric Gauss-Seidel smoothing via PETSc's smoothed aggregation-based GAMG implementation.

Smoothed aggregation is an approach to AMG which involves grouping fine-grid degrees of freedom into coarser grid spaces, targeting error components that are less effectively handled by smoothing alone.
Once aggregates are determined, smoothing steps are applied to enhance the effectiveness of the prolongation operators.
We choose a single smoothing step to be applied in the construction of our prolongation operators on each level.
GAMG provides a wide range of tuning options, including choices for the smoother, the number of pre- and post-smoothing iterations, and thresholds for controlling coarsening rates and aggregation construction.
An exhaustive optimization of these parameters is beyond the scope of this current work.

Since the PETSc solver itself accounts only for a small fraction of the total computing time, the solvers are parallelized using only MPI and no GPU-offloading.
$2 N_\varphi$ equations with a total of $3N_\varphi N_n$ unknowns have to be solved with $N_{MPI}$ MPI ranks.
The available MPI ranks are divided into $N_\varphi$ groups with $N_{plane}=N_{MPI}/N_\varphi$ MPI ranks, one group for each plane.
Since each plane group may still be too large to efficiently solve a single field equation, they are subdivided into smaller solver groups with a number of MPI ranks set such that $N_n/N_{solver} \gtrsim$ 5,000, which is chosen as a heuristic strong scaling limit of the iterative PETSc solvers used in XGC.
If a plane group can accommodate more than one solver group, which is often true, two or more field equations can in principle be solved simultaneously.
For example, a simulation of ITER with $N_n=1.5\cdot 10^6$ and $N_\varphi=32$ is typically run on at least 2,000 Frontier \cite{Chang2024} compute nodes with 8 MPI ranks per node, i.e., $N_{plane}>500$, which is close to twice the target for the number of ranks in the solver group ($N_{solver} \lesssim N_n/5,000=300$).

\begin{figure}
 \centering
 \includegraphics[width=0.95\textwidth]{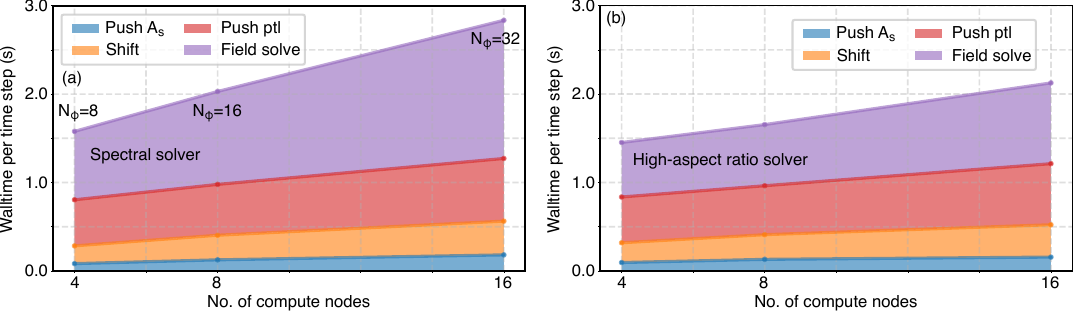}
 \caption{Execution time per time step with the new spectral solver (a), and the approximate solver using the high aspect ratio approximation (b) for the internal kink mode benchmark discussed in Sec. \ref{subsec:veri_kink}. The number of poloidal planes $N_\varphi$ (i.e., toroidal resolution) and particles (100 ptl/cell, 32,000 vertices per plane) scale with the number of compute nodes.}
 \label{fig:solver_weak_scaling}
\end{figure}

In contrast to the PETSc solver, the Fourier filters required for the solver algorithm described in Sec. \ref{subsec:high-n_solver}, are performance critical due to the large amount of MPI communication they require.
The filter is applied to each flux surface independently, providing a natural parallelization.
However, XGC’s mesh vertices are not guaranteed to be in flux-surface order.
This could be dealt with using all-to-all communication to ensure that the input values are available to the assigned MPI process.
But since the amount of data that must be communicated to each MPI process scales linearly with toroidal resolution (i.e. the number of poloidal planes), the approach using all-to-all communication proved to scale poorly.
To avoid the expensive communication operation, the data is temporarily reordered so that the values on each flux surface are contiguous in memory.
The data is then distributed across all ranks assigned to a poloidal plane, each of which need only exchange small amounts of data with other planes.
As a result, total time spent in the filter is reduced to acceptable levels even for high-resolution simulations.
Using the example of the internal kink case discussed in Sec. \ref{subsec:veri_kink}, Figs. \ref{fig:solver_weak_scaling} (a) and (b) indicate an acceptable increase of the walltime per time step due to the spectral solver.
The plots show a weak scaling test performed on the GPU partition of Perlmutter at NERSC in which the number of poloidal planes $N_\varphi$ (i.e., toroidal resolution) and particles (100 ptl/vertex, 32,000 vertices per plane) scale with the number of compute nodes.
The time marked for the field solve in the plots includes all operations required for solving the field equations, i.e., charge/current density scatter, filtering and post-processing.
Most of the additional computing time from the spectral solver (compared to the high aspect ratio solver) is due to the additional pre- and post-processing needed by the spectral solver, not due to the solver itself.
Note that this $\delta f$ case has a particularly light particle workload with roughly 6.5 million markers per species per GPU.
This is the reason why the field solve makes up for a relatively large fraction of the overall execution time.
In larger $\delta f$ cases (e.g., for ITER) and in total-$f$ simulations ($\sim$ 10,000 ptl/vertex, with Fokker-Planck collisions), the field solver accounts for a much smaller part of the execution time.

\section{Error Analysis}\label{sec:error_analysis}
%
%
For a quantitative analysis of the deviation between the solutions of the approximated (Eqs. \eqref{eq:a_op_approx} and \eqref{eq:b_op_approx}) and accurate (\eqref{eq:pol_op_spectral_a} and \eqref{eq:pol_op_spectral_b}) field solvers, we define a test function for the right-hand-side, and a normalized error measure.
For the purpose of this error analysis, we solve the field equation with $M_{ad}=0$ because this term (e.g., the adiabatic response in case of the gyrokinetic Poisson equation) can dominate the left hand side of the equation. 
We evaluate the error measure for two magnetic field geometries, one in which no meaningful error is expected, and one for which finite aspect ratio effects are expected to be significant.

The test function is defined as a single 2D Fourier mode
\begin{equation}\label{eq:error_test_function}
    \rho = \rho_0 \exp\left[- \left( \frac{\psi-\psi_c}{w} \right)^6 \right] \cos\left( n\varphi + m\theta^\ast \right)
\end{equation}
with $\psi_c=0.5$ and $w=0.38$.
The normalized error measure is defined with the $(n,m)$ Fourier coefficients of the solution of the field equation $\phi_{n,m}$ as
\begin{equation}\label{eq:error_measure}
    E(n,m) = \sqrt{\frac{\left| \phi_{n,m}^{(approx)} - \phi_{n,m}^{(spec)} \right|^2}{\left| \phi_{n,m}^{(spec)} \right|^2}}.
\end{equation}
For this analysis, we retain only modes that are resolved with at least eight mesh vertices per wavelength.
The error for all other modes is set to zero in the plots shown in this section.

The first test cases we use is the circular, Cyclone-like \cite{Dimits2000} case used in Ref. \onlinecite{Sturdevant2021} with major radius $R_0=2.8$ m, minor radius $a=1$ m (inverse aspect ratio $\varepsilon=0.357$), central magnetic field $B_0=0.236$ T, temperature $T_0=T(r/a=0.5)=2.14$ keV ($\rho^\ast\approx 1/50$), density $n_0=n(r/a=0.5)=3.23\cdot 10^{17} \,\mathrm{m}^{-3}$, and safety factor $q=0.86-0.16 (r/a) + 2.52 (r/a)^2$.
The second test case is NSTX discharge 132588 with $R_0=1.05$ m, $a=0.62$ m ($\varepsilon=0.59$), $B_0=0.37$ T, $T_0=1.25$ keV ($\rho^\ast\approx 1/50$), $n_0=6.99\cdot 10^{19} \,\mathrm{m}^{-3}$, and $q_{95}=8.0$.

\begin{figure}
    \centering
    \includegraphics[width=0.95\textwidth]{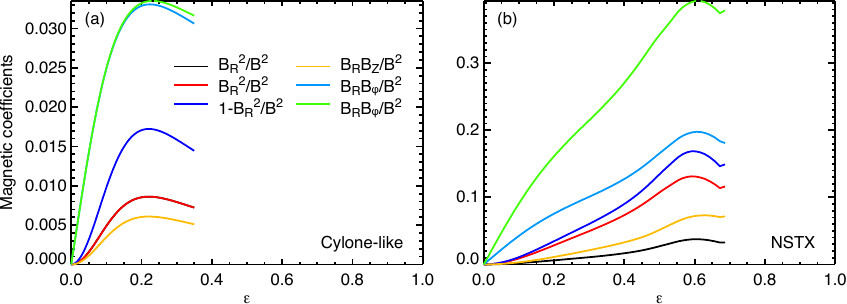}
    \caption{Magnetic field coefficients in Eqs. \eqref{eq:pol_op_spectral_a} and \eqref{eq:pol_op_spectral_b} as functions of the inverse aspect ratio for (a) a Cyclone-like magnetic field geometry and (b) NSTX discharge 132588.}
    \label{fig:error_mag_field_coeffs}
\end{figure}

The values of the magnetic field coefficients in Eqs. \eqref{eq:pol_op_spectral_a} and \eqref{eq:pol_op_spectral_b} for the two cases are shown as functions of the inverse aspect ratio in Fig. \ref{fig:error_mag_field_coeffs}.
The magnetic coefficients in case of the Cyclone-like geometry are all very small (below 3\%).
So the large-aspect approximation is expected to be fairly accurate.
In case of the NSTX discharge, however, the largest of the coefficients approaches 40\%, and the others range between 5 and 20\%.
Notably, the two largest coefficients are the ones that couple the poloidal sine and cosine components of the fields.
Therefore, a significant deviation between the large-aspect ratio approximated and the accurate field equations can be expected.

\begin{figure}
    \centering
    \includegraphics[width=0.95\textwidth]{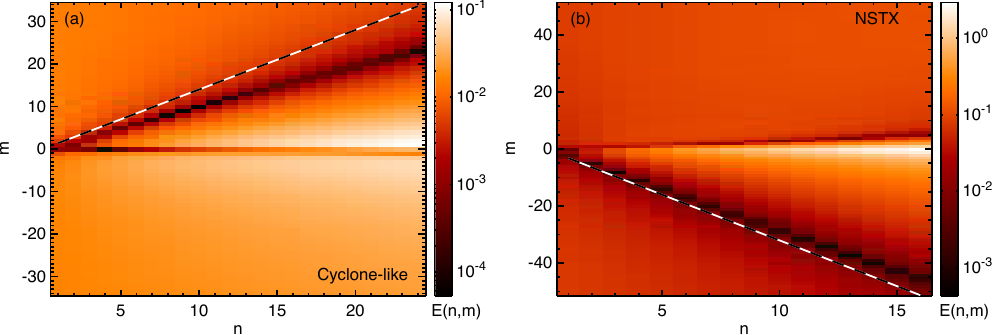}
    \caption{Normalized difference $E(n,m)$ from Eq. \eqref{eq:error_measure} between the accurate and the large aspect ratio approximated field equations for the test function $\rho$ in Eq. \eqref{eq:error_test_function}: (a) Cyclone-like magnetic geometry at $\varepsilon=0.18$, (b) NSTX discharge 132588 at $\varepsilon=0.41$.}
    \label{fig:error_2D}
\end{figure}

\begin{figure}
    \centering
    \includegraphics[width=0.5\textwidth]{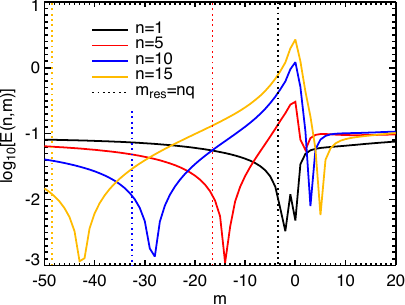}
    \caption{Cross sections of the error measure of the approximated field equation at various toroidal mode numbers for NSTX discharge 132588 at $\varepsilon=0.41$ and $q=3.2$.
    The dotted vertical lines indicate the resonant mode $m=nq$ for each $n$.}
    \label{fig:error_1D}
\end{figure}

The error measure $E(n,m)$ is shown in Fig. \ref{fig:error_2D}  (a) for the Cyclone-like case at $\varepsilon=0.18$ and $q=1.4$, and (b) for NSTX discharge 132588 at $\varepsilon=41$ and $q=3.2$.
The black-white dashed line indicates the resonant modes $m=n q$.
The two most obvious observations in both cases are: i) a line with minimal error and poloidal mode numbers that are proportional to the toroidal mode number, but not exactly resonant; and ii) a line with maximal error at low poloidal mode number that increases with the toroidal mode number.
For field aligned modes, the difference between the approximated and exact field equation is a few percent at most, which is well below the estimate in Eq. \eqref{eq:k_perp_error}.
This is illustrated in Fig. \ref{fig:error_1D} for the NSTX discharge, which exhibits large errors around $m=0$, but reduced errors close to $m=nq$.
But in case of the NSTX discharge, the relative difference between the approximated and the exact field equation is much larger, up to order unity, for low poloidal mode numbers.

The relevant question with regard to the applicability of the approximated field equation is the error within reasonable bounds of $k_\parallel \approx N_\parallel/R_0$, where $|N_\parallel| = |m/q-n| \leq N_\parallel^{(max)}$ is the number of mode periods along the magnetic field over one toroidal circuit.
For XGC, we consider $N_\varphi = 8 N_\parallel^{(max)}$ as good numerical resolution.
In many practical cases $N_\varphi\sim32$.
Therefore, $N_\parallel^{(max)}=4$ is a good practical limit.
\begin{figure}
    \centering
    \includegraphics[width=0.95\linewidth,interpolate=false]{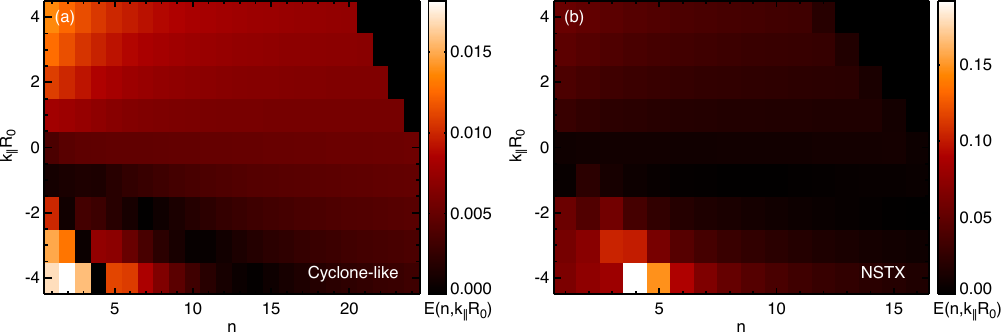}
    \caption{Relative difference between the approximated and exact field equation vs. toroidal mode number $n$ and parallel wavenumber $k_\parallel$ for (a) Cyclone-like geometry at $\varepsilon=0.18$, and (b) NSTX discharge 132588 at $\varepsilon=0.41$.}
    \label{fig:error_kpara}
\end{figure}
Figure \ref{fig:error_kpara} shows the relative difference $E(n,k_\parallel)$ between the approximated and accurate field equations for the Cyclone-like case at $\varepsilon = 0.18$ [Fig. \ref{fig:error_kpara} (a)], and for NSTX discharge 132588 at $\varepsilon=0.41$ [Fig. \ref{fig:error_kpara} (b)].
In both cases, the difference peaks at low toroidal mode numbers and negative $k_\parallel$ (corresponding to low poloidal mode number).
The error decreases with increasing toroidal mode number and remains at an acceptable level of a few percent over the entire practical range of $k_\parallel$.
Figure \ref{fig:error_n_psi} shows that this is true at small and finite aspect ratio for NSTX 132588, and that the error approaches the high toroidal mode number limit in Eq. \eqref{eq:k_perp_error}.
\begin{figure}
    \centering
    \includegraphics{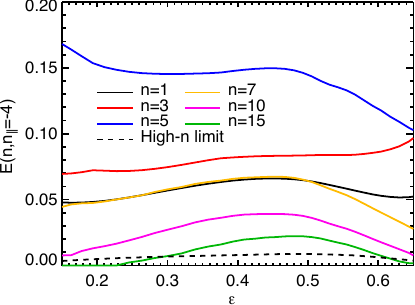}
    \caption{Relative difference between the solutions of the approximated and the accurate field equation for NSTX discharge 132588 and $k_\parallel R_0=-4$ as function of the normalized poloidal flux $\psi_N$ for different toroidal mode numbers. The dashed line shows the limit of the error for high toroidal mode numbers, Eq. \eqref{eq:k_perp_error}.}
    \label{fig:error_n_psi}
\end{figure}
%

In conclusion, the observations of the error analysis in this section justify the approach of using the accurate field equation only for lower toroidal mode numbers, while higher toroidal mode numbers are solved with the approximated field equation.
Using the approximated field equation alone for all toroidal mode numbers in higher aspect ratio geometries is also acceptable.


\section{Verification}\label{sec:verification}
For physics verification of the new field solver, we exercise several basic electromagnetic physics cases.
In Sec. \ref{subsec:veri_alfven}, we verify the dispersion relation of shear-Alfv\'{e}n modes against an analytical formula, and XGC with the approximate field equations and implicit time integration \cite{Sturdevant2021}.
In Secs. \ref{subsec:veri_kink} and \ref{subsec:veri_tearing}, we study the $(n,m)=(1,1)$ internal kink mode with XGC and NIMROD \cite{Sovinec2004, Held2015} (comparing to results from GTS \cite{Startsev2024}), and the collisionless tearing mode with XGC and ORB5 \cite{Lanti2020}.
In Sec. \ref{subsec:veri_turb}, we study nonlinear turbulence at low toroidal mode numbers in the Cyclone-like case introduced in Sec. \ref{sec:error_analysis}.

\subsection{Shear-Alfv\'{e}n mode verification}\label{subsec:veri_alfven}
As a basic physics benchmark, we study the dispersion of the shear-Alfv\'{e}n wave.
For the theoretical prediction of the real frequency and damping rate for our shear-Alfv\'{e}n benchmark problem, we begin with the following dispersion relation, which was originally derived for shear-Alfv\'{e}n wave propagation in slab geometry \cite{Sturdevant2022}
\begin{equation}\label{eq:alfven_dispersion}
    1 - \frac{1}{2 k_{\perp}^2 \rho_s^2} \left( 1 - \left( \frac{\omega}{k_\parallel v_e} \right)^2 \frac{m_i}{2 m_e} \beta_e \right) Z'\left( \frac{\omega}{\sqrt{2} k_\parallel v_e} \right) = 0.
\end{equation}
%
Here $Z$ is the plasma dispersion function of Fried and Conte \cite{Fried1961}, and $\beta_e=2\mu_0 n_0 k_B T_0/B_0^2$.
Taking an initial density perturbation of the form
\begin{equation}\label{eq:veri_alfven_init_pert}
\delta n = \epsilon \sin{(k_r r)} \cos(n \varphi + m \theta^*),
\end{equation}
we approximate $k_{\perp}$ and $k_\parallel$ to lowest order in $\varepsilon$ as
\begin{align}
k_\parallel & \approx \frac{1}{R_0} \left( \frac{m}{q_0}-n \right), \notag\\
k_\perp & \approx \sqrt{\left( \frac{m}{r_0} \right)^2 + k_r^2 }. \label{eq:alfven_veri_k_definition}
\end{align}
We take $r_0$ as the coordinate of the surface where the measurement is taken.

We compare two versions of XGC against Eq. \eqref{eq:alfven_dispersion} using a reduced $\delta f$ algorithm: the standard electromagnetic version with an explicit time integrator \cite{Hager2022} and the accurate field equations discussed in this article, and another with implicit time integrator \cite{Sturdevant2021} that employs the approximate field equations.
The XGC simulations are exercised in a setup used also in Ref. \onlinecite{Sturdevant2021}: in an analytic magnetic field geometry with concentric circular flux-surfaces, $R_0=50$ m and $R_0=4$ m, $a=1$ m (i.e., $\varepsilon=0.02$, and $\varepsilon=0.25$), $B_0=0.228$ T, and uniform safety factor $q=2$, density $n_0=3.125 \cdot 10^{17} \,\mathrm{m^{-3}}$, and temperature $T_0=2$ keV ($\beta_e=4.8 \cdot 10{-3}$, $a/\rho_i=1/35$).
We use deuterium ions ($m_i=m_D$) and slightly heavier electrons, $m_i/m_e=1836$.
The mesh resolution is 0.35 $\rho_i$ in the radial and on average 0.25 $\rho_i$ in the poloidal direction with $N_\varphi=32$ and $N_w=1$.
This is sufficient to comfortably resolve $|n_\parallel|=|m/q-n| \leq 4$.
For the initial condition, Eq. \eqref{eq:veri_alfven_init_pert}, we chose $n=2$, and $k_r=12.7 \,\mathrm{m^{-1}}$, and we scan the Alfv\'{e}n frequency and damping rate over the poloidal mode number $m$, which is equivalent to scanning $k_\parallel$ because $n$ is fixed.
The time steps used in the simulation with explicit time integrator are $\Delta t=4.4 \cdot 10^{-7}$ s ($\varepsilon=0.02$) $\Delta t=3.5 \cdot 10^{-8}$ s ($\varepsilon=0.25$), or roughly $0.055 \tau_A$, where $\tau_A = R_0/v_A = (R_0/B_0)(\mu_0 n_0 m_i)^{1/2}$ is the Alfv\'{e}n time.
The corresponding parallel mode number range is $-3 \leq n_\parallel \leq 3$.
Frequencies and damping rates are measured at $r=0.35$ m.

\begin{figure}
    \centering
    \includegraphics[width=0.97\textwidth]{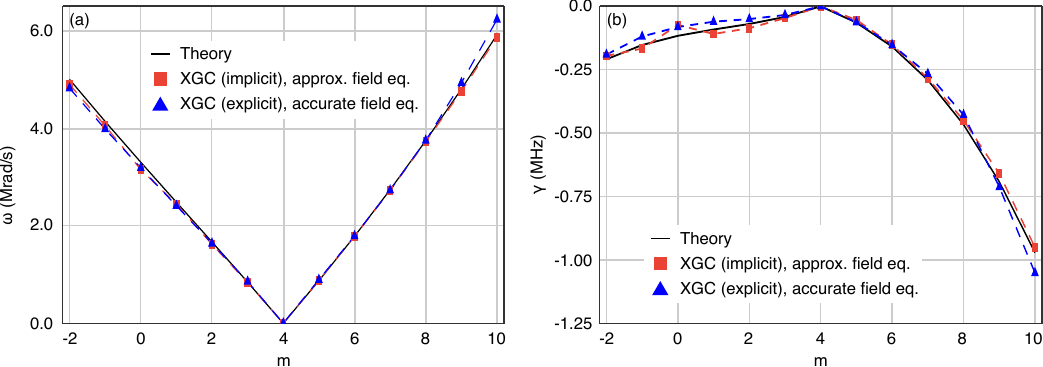}
    \caption{(a) Frequency, and (b) damping rate of the $n=2$ shear-Alfv\'{e}n wave in a magnetic geometry with concentric circular flux-surfaces, inverse aspect ratio $\varepsilon=0.25$, and safety factor $q=2$.}
    \label{fig:alfven_dispersion}
\end{figure}

The mode frequencies and damping rates for the case with $\varepsilon=0.25$ are shown in Fig. \ref{fig:alfven_dispersion}.
The results of the XGC simulations closely match each other and the theoretical prediction of Eq. \eqref{eq:alfven_dispersion} at both inverse aspect ratios.

\subsection{Internal kink mode verification}\label{subsec:veri_kink}
The next verification of MHD-type modes is a benchmark of XGC against the MHD code NIMROD \cite{Sovinec2004} with a drift-kinetic closure \cite{Held2015}, and GTS \cite{Startsev2024} for the $(n,m)=(1,1)$ internal kink mode with hydrogen ions and an ion-electron mass ratio of $m_e/m_i=5.4 \cdot 10^{-3}$.
The numerical setup is identical to the one used in Ref. \onlinecite{Startsev2024} and similar to Ref. \onlinecite{Mishchenko2019}, concentric circular flux-surfaces with $R_0=10$ m, $a=1$ m, $B_0=1$ T, $T_i=T_e=2.957$ keV, and $\rho^\ast=1/180$.
The safety factor profile is $q=0.8[1+(r/a)^2]$ with the $(n,m)=(1,1)$ resonant surface at $r/a=0.5$.
The density profile is defined as
\begin{equation}
    n(\tilde{s}) = n_0 \exp\left\lbrace -\delta_n \kappa_n \mathrm{tanh}\left[ \left(r/a-s_0 \right)/\delta_n \right]\right\rbrace,
\end{equation}
where $n_0=2^i \cdot 8.40 \cdot 10^{18} \,\mathrm{m^{-3}}$, $\delta_n=0.2$, $\kappa_n=3$, and $s_0=0.5$, and $i=$0, $-1$ and $-2$ so that $\beta_e$ is 1.0\%, 0.5\% and 0.25\%.
XGC employs the electromagnetic $\delta f$ method from Ref. \cite{Cole2021}, which is also implemented in EUTERPE \cite{Kleiber_2016}, ORB5 \cite{Mishchenko2019}, and GTS \cite{Startsev2024}.
For the free energy drive from the background current density gradient, XGC uses the current density at the outer midplane, $j_\parallel = (\hat{\boldsymbol{b}} \cdot \nabla \times \boldsymbol{B}/\mu_0) (\theta=0)$, similar to ORB5.
(GTS uses the local current density gradient\cite{Startsev2025_private}.)
XGC is set up with a smoothly varying radial resolution of $\Delta r \approx \rho_i$ at $r/a=0$ and $r/a=1$, and $\Delta r \approx \rho_i/2$ at $0.4 < r/a < 0.55$.
The poloidal resolution is $3\rho_i < \Delta l_\theta \approx 2\Delta < 5\rho_i$ at $0.3<\psi_N<0.7$, $N_\varphi=8$, and $N_w=1$, which corresponds to a resolution of $n\leq 64$ and $k_\parallel R_0 \leq 1$.
The time step in the XGC simulation is $\Delta t=2.2\omega_{ci} = 3.6\cdot 10^{-2}\, \mathrm{c_s/L_n}$, where $\omega_{ci}$ is the ion Larmor frequency, $c_s=(e T_e/m_i)$ is the sound speed, and $L_n=1/\kappa_n$ is the density gradient scale length at $r/a=s_0$.
We apply a magnetic field-aligned mode filter that limits the spectrum to $n=1$ and $|k_\parallel R_0| \sim |m/q-n| \leq 1$, which is effectively equivalent to the constraint $0\leq m \leq 2$ used in Ref. \onlinecite{Startsev2024}.

For the NIMROD results, a 32$\times$48 finite element mesh with biquadratic basis functions was employed.
(Bicubic basis functions changed the growth rates by only a few percent.)
The time step for the ideal MHD and kinetic scans was $10^{-7}$ s, and for the collisionless two-fluid scan it was $2\cdot10^{-8}$ s. 
For the kinetic scan, 30 degrees of freedom (3 speed points and 10 pitch-angle basis functions) were used to resolve the electron and ion distribution functions, with the kinetic physics coupling to NIMROD's fluid model through the pressure tensor closures.

\begin{figure}
    \centering
    \includegraphics[width=0.97\textwidth]{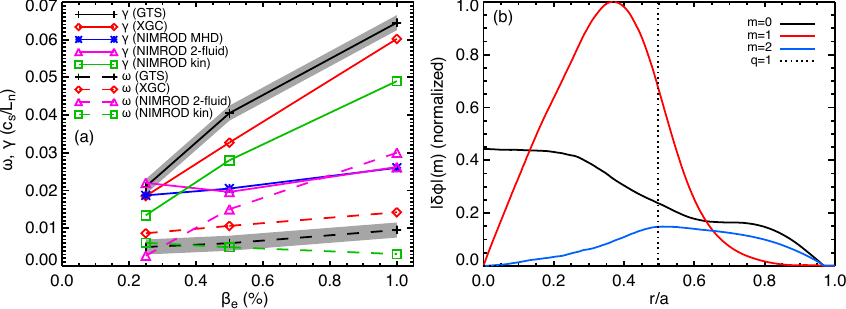}
    \caption{(a) Growth rate and frequency of the $(n,m)=(1,1)$ internal kink mode in XGC, NIMROD (MHD, 2-fluid, and MHD+kinetic closure) compared to GTS \cite{Startsev2024}. (b) Radial envelope of the Fourier modes $m=1$, 2, and 3 of the electrostatic potential (compare to Fig. 7 in Ref. \onlinecite{Startsev2024}).}
    \label{fig:kink_gamma_comparison}
\end{figure}

The growth rates and frequencies of the $(n,m)=(1,1)$ kink mode from XGC, GTS and NIMROD are shown in Fig. \ref{fig:kink_gamma_comparison} (a).
For GTS (solid and dashed black lines with crosses), we included the growth rate from the eigenmode solver, and the frequency from the GTS simulations in Ref. \onlinecite{Startsev2024}.
Since we read the values from Fig. 7 in Ref. \onlinecite{Startsev2024}, we mark an error range of $2\cdot 10^{-3}\,\mathrm{c_s/L_n}$ in grey.
The XGC results (solid and dashed red lines with diamonds) agree reasonably well with the magnitude and $\beta_e$ dependence of the GTS results.
The frequencies from XGC are between 6\% ($\beta_e=1.0\%$) and 19\% ($\beta_e=0.5\%$) below the values from the GTS eigenmode solver.
The mode frequencies observed in XGC are slightly higher than the corresponding GTS values.

For NIMROD, we show results from three different scans, one using the ideal MHD equations (solid blue line with stars), one using collisionless 2-fluid equations (solid and dashed magenta lines with triangles), and one using MHD with a kinetic closure for the thermal pressure in the flow equation (solid and dashed green lines with squares).
The NIMROD results illustrate the important physics added by the two-fluid model and kinetic closure.
The MHD and two-fluid results exhibit roughly the same growth rates.
But while the frequency of the kink mode in the ideal MHD model is zero, the 2-fluid model yields a finite frequency (albeit different from the kinetic frequency) due to the added diamagnetic flow effects.
The corrections in NIMROD from the kinetic closure for the pressure tensor in the flow equation introduce trapped particle corrections that lead to order unity corrections of the growth rate, and bring the NIMROD results for both the growth rate and the frequency close to the XGC and GTS results.

Note that large corrections due to trapped particles have been found before with a destabilizing effect due to energetic particles \cite{Chen1984}, and a stabilizing effect due to thermal particles \cite{Antonsen1993}.
The destabilizing effect with energetic particles was found to be due to a resonance with the hot ion toroidal precession frequency $\omega_d$, while the stabilizing effect with thermal particles was found for mode frequencies and growth rates much smaller than the bounce frequency $\omega_b$ and the precession frequency.
The present case is somewhat different in that $\gamma > \omega \sim \omega_b$, i.e., the kink mode can resonate with the trapped ions on the bounce time scale.

The normalized radial mode structure of the electrostatic poloidal for mode numbers $m=$0, 1, and 2 is shown in Fig. \ref{fig:kink_gamma_comparison} (b).
The curves for $m=1$ (red) and $m=2$ (blue) agree well with the curves shown in Fig. 7 (b) of Ref. \onlinecite{Startsev2024}.
Note that unlike Ref. \onlinecite{Startsev2024}, we show the absolute value of the Fourier coefficient $|\phi|(r/a,m)$ instead of the real and imaginary parts because their respective magnitude depends on the phase of the kink mode due to its finite frequency.
%

\begin{figure}
    \centering
    \includegraphics[width=0.97\textwidth]{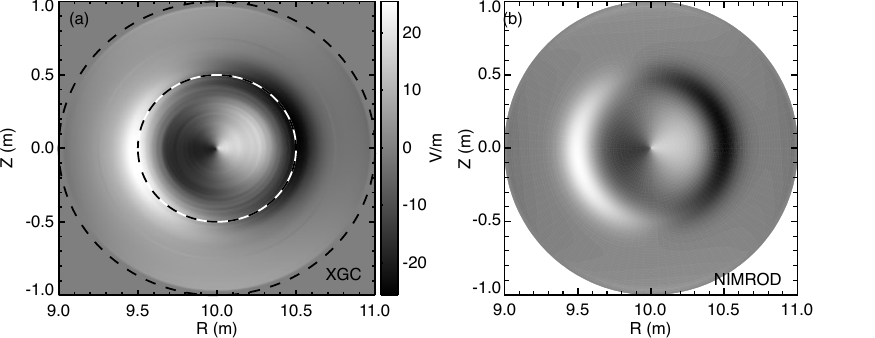}
    \caption{Mode structure of the $(n,m)=(1,1)$ component of the radial electric field $E_r$ of the kink mode in (a) XGC and (b) NIMROD.}
    \label{fig:kink_mode_structure}
\end{figure}

Figure \ref{fig:kink_mode_structure} shows radial-poloidal cross-sections of the radial electric field $E_r = \nabla\psi/|\nabla\psi| \cdot \boldsymbol{E}$ from (a) XGC, and (b) NIMROD.
The mode structures of the internal kink mode observed with XGC and NIMROD is almost identical.
Both are dominated by the contribution from the resonant mode $m=1$, exhibit a maximum of $E_r$ at the resonant surface ($r/a=0.5$), and a change of the sign of $E_r$ due to the local maximum of the electrostatic potential shown in Fig. \ref{fig:kink_gamma_comparison} (b).
%

With all three codes, XGC, NIMROD and GTS clearly finding the same mode in terms of the mode structure [Figs. \ref{fig:kink_gamma_comparison} (b) and \ref{fig:kink_mode_structure}] and $\beta_e$ dependence of the growth rates, the benchmark of the internal kink mode is generally successful.
The scatter in the growth rates and frequencies [Fig. \ref{fig:kink_gamma_comparison} (a)] seems acceptable given the differences between XGC, NIMROD and GTS.
NIMROD is a Eulerian fluid code with optional drift-kinetic closure, which makes it quite different from XGC and GTS.
In addition, the test setup uses an idealized but strictly inconsistent magnetic field geometry (concentric circular flux-surfaces with a finite pressure gradient).
This is not an issue in $\delta f$ gyrokinetic simulations because the physics of the MHD equilibrium have been ordered out of the equations.
But for NIMROD, this test case required us to override the calculation of the MHD equilibrium magnetic field that is normally carried out in MHD codes.
And although XGC and GTS nominally use the same electromagnetic gyrokinetic method and are both PIC codes, their implementation is very different, with GTS employing flux-coordinates, and XGC employing cylindrical coordinates.
Other subtle differences may be in the Fourier filters applied in both XGC and GTS (toroidal and poloidal) and in NIMROD (toroidal only), and in the algorithms used to determine the growth rates and frequencies.

\subsection{Collisionless tearing mode verification}\label{subsec:veri_tearing}
In this section, we verify the XGC against ORB5 for a collisionless tearing mode with $(n,m)=(1,2)$ (similar to Ref. \onlinecite{Mishchenko2022}) using a reduced $\delta f$ formalism.
The background magnetic field has circular flux-surfaces with $R_0=10$ m, $a=1$ m, $B_0=1$ T, and uniform density $n_0=5.19 \cdot 10^{17} \,\mathrm{m^-3}$ and $T_0=9.581$ keV, $m_i=m_p$, $m_i/m_e=200$, $\rho^\ast=1/100$, and $\beta_e=0.2\%$.
The safety factor profile is given by
\begin{equation}\label{eq:tearing_veri_q}
    q = q_a\,\frac{\left( r/a \right)^2}{1-\left[1-\left( r/a \right)^2\right]^2},
\end{equation}
with $q_a=3.5$ so that the $q=2$ surface is at $r/a=0.5$.
In this configuration, the $(n,m)=(1,2)$ tearing mode is driven entirely by the equilibrium current density profile, which is taken to be the equilibrium current density along the outer midplane, $j_\parallel = (\hat{\boldsymbol{b}} \cdot \nabla \times \boldsymbol{B}/\mu_0) (\theta=0)$.

The simulation mesh used by XGC has a smoothly varying radial resolution between $0.3 \rho_i$ at $0.4<r/a<0.6$ and $\rho_i$ at $r/a<0.2$ and $r/a>0.8$.
The poloidal mesh resolution varies between $0.6 \rho_i$ and $\rho_i$ (governed by the field-alignment constraint) with $N_\varphi=16$ and $N_w=1$.
XGC uses a time step of $\Delta t \omega_{ci}=3.1$, and 100 marker particles per mesh vertex and species.

\begin{figure}
    \centering
    \includegraphics[width=0.97\textwidth]{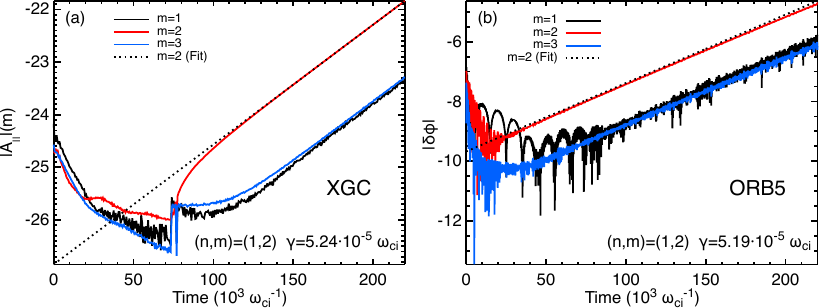}
    \caption{Comparison of the growth rate of the collisionless tearing mode between XGC and ORB5.}
    \label{fig:tearing_growthrate}
\end{figure}

The ORB5 simulation is nonlinear with a single toroidal harmonic $n=1$.
In contrast, Ref. \onlinecite{Mishchenko2022} considers $-30 \le n  \le 30$.
Only the linear phase is shown in Fig. \ref{fig:tearing_growthrate} (b).
The radial grid resolution is $N_s = 307$, the poloidal resolution $N_{\theta} = 32$, and the toroidal resolution $n_{\varphi}=16$ grid points.
The time step is $\Delta t = 5.0 \omega_{ci}^{-1}$.
A total of 200 million ion and 400 million electron markers is used corresponding to about 1,300 and 2,500 markers per mesh vertex, respectively. (This is a nonlinear simulation requiring good statistics.)

The growth rates of the $(1,2)$ collisionless tearing mode obtained with XGC ($\gamma=5.24 \cdot 10^{-5} \omega_{ci}$) and ORB5 ($\gamma=5.19 \cdot 10^{-5} \omega_{ci}$) shown in Fig. \ref{fig:tearing_growthrate} match almost perfectly with a difference of only about 1\%.
Both codes find $(n,m)=(1,2)$ to be the dominant mode with by far with a clear separation in amplitude to the modes $(1,1)$ and $(1,3)$.

\begin{figure}
    \centering
    \includegraphics[width=0.97\textwidth]{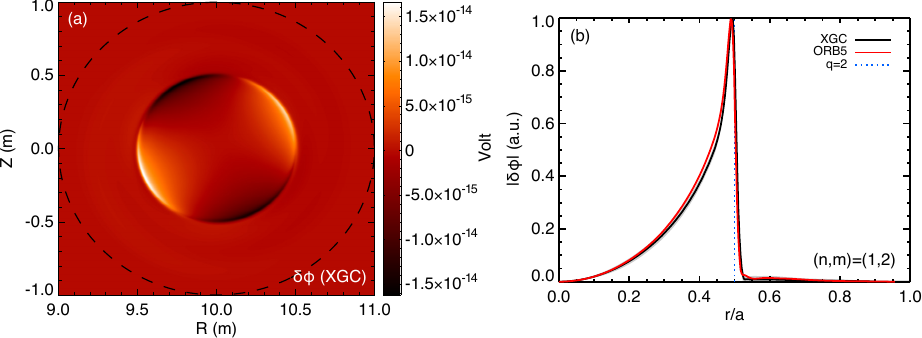}
    \caption{Mode structure of the collisionless $(n,m)=(1,2)$ tearing mode. (a) Electrostatic potential on a radial-poloidal cross section. (b) Normalized amplitude of the $(n,m)=(1,2)$ Fourier coefficient of the electrostatic potential. The dotted line indicates the $q=2$ resonant surface.}
    \label{fig:tearing_mode_struc}
\end{figure}

The mode structure of the electrostatic potential perturbation is shown in Fig. \ref{fig:tearing_mode_struc} (a).
It exhibits a clear $m=2$ mode with a sharp cutoff on the resonant $q=2$ surface.
The time-averaged ($150\cdot 10^3 \omega_{ci}<t<220\cdot 10^3 \omega_{ci}$) normalized radial envelope of the electrostatic potential perturbation shown in Fig. \ref{fig:tearing_mode_struc} (b) closely tracks the one obtained with ORB5.
The uncertainty of the radial mode envelope in XGC is defined as three times the standard deviation of the envelope's time evolution and is indicated by the gray shaded area.

\subsection{Nonlinear turbulence verification}\label{subsec:veri_turb}
%
We verify nonlinear turbulence behavior with the new field solver by comparing to the old (large aspect ratio) solver in a case in which the difference between the two can be expected to be small.
Our simulations use the Cyclone-like case introduced in Sec. \ref{sec:error_analysis} with the density and temperature profiles defined by their logarithmic gradient \cite{Sturdevant2021}
\begin{equation}
    k_A = -R_0\,\frac{\partial(\ln A)}{\partial r} = k_A(r_0)\exp\left[-\left((r-r_0)/(w_A a)\right)^6\right],
\end{equation}
where $A$ can be the density or temperature, and $r_0=a/2$, $n(r_0)=3.23\cdot 10^{17}\,\mathrm{m^{-3}}$, $T(r_0)=2.14$ keV, $k_n(r_0)=2.22$, $k_T(r_0)=6.92$, and $w_n(r_0)=w_T(r_0)=0.25$.
The main ion species is hydrogen ($m_i=m_p$), and $m_e$ the real electron mass.
This corresponds to $\rho^\ast\sim 1/50$, and $\beta_e=0.5\%$.

The simulations with the two field solvers are carried out using the reduced $\delta f$ formalism.
The configuration space mesh used for this study has 28,439 vertices on each radial-poloidal plane, corresponding to a radial and poloidal resolution of $\Delta r/\rho_i=\Delta l_\theta/\rho_i=0.35$, with $N_\varphi=24$ and $N_w=1$.
A toroidal-poloidal Fourier filter is applied that limits the parallel wave number to $k_\parallel R_0 \leq 3$, and to $0<n<12$.
The time step of $\Delta t \omega_{ci}=0.1$ is sufficient to resolve the fastest Alfv\'{e}n wave allowed by the Fourier filter, $\omega_A=3 v_A/R_0$ with $v_A=R_0/(\mu_0 n(r_0) m_i)^{1/2}$, with 12 time steps.

\begin{figure}
    \centering
    \includegraphics[width=0.97\textwidth]{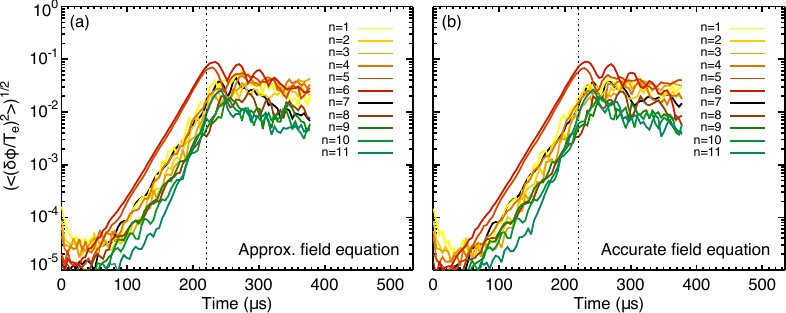}
    \caption{Amplitude of toroidal mode numbers $0<n<12$ of the electrostatic potential at $r/a=0.45$ in a (nonlinear) electromagnetic reduced $\delta f$ XGC simulation of the Cyclone-like case introduced in Sec. \ref{sec:error_analysis} using (a) the approximated field equations, and (b) the accurate field equations.}
    \label{fig:ecbc_turb_verification}
\end{figure}

The simulations are run through the exponential growth phase into the saturated turbulence phase for a total time of $382 \,\mathrm{\mu s}$.
The time evolution of the amplitudes of the toroidal mode numbers $0<n<12$ of the electrostatic potential at $r/a=0.45$ are shown in Fig. \ref{fig:ecbc_turb_verification} (a) for the simulation with the approximate field equations, and (b) with the accurate field equations.
In the exponential growth phase between $t=0\,\mathrm{\mu s}$ and $t\sim 220\,\mathrm{\mu s}$, the mode with the highest amplitude is $n=6$ in both simulations, closely followed by $n=5$.
Turbulence saturation sets in at $t\sim 220\,\mathrm{\mu s}$ when the fastest growing mode reaches an amplitude of $\delta\phi/T_e\sim 10\%$.
In the saturated turbulence phase, the mode spectrum is dominated by $n\leq 6$.
As expected based on the error analysis in Sec. \ref{sec:error_analysis}, the simulations with the approximate and accurate field equations show excellent agreement.

\section{Conclusions}\label{sec:conclusions}
To improve the accuracy of XGC for reduced MHD-type instabilities with low toroidal mode numbers, we developed, implemented, and verified a new field solver for the gyrokinetic Poisson equation and Amp\'{e}re's law.
This improvement was necessary because XGC's standard field solver, similar to other gyrokinetic codes \cite{Michels2021}, employs a large aspect ratio approximation to reduce the numerical complexity of the field equation.

A simple error estimate in Sec. \ref{subsec:polarization_error_estimate} showed that this approximation is justified for micro-turbulence in tokamak edge plasma, where typically both the toroidal mode number and the safety factor are relatively large.
But reduced-MHD modes such as tearing, kink, or peeling-ballooning modes have low to moderate toroidal mode numbers, and some may be located in the core at low safety factors.
In this regime, the approximate field solver can incur significant errors.

For more accurate simulations of reduced-MHD modes, we introduced in Sec. \ref{sec:derivation} a toroidally spectral discretization of the gyrokinetic Poisson equation and Amp\'{e}re's law without using the large aspect ratio approximation.
Since the spectral discretization decomposes the 3D field equations into a set of 2D equations on the radial-poloidal plane similar to the large aspect ratio approximation, our approach had the big advantage that the existing solver infrastructure of XGC could be reused to a large extent.
For combined simulations of the interaction of reduced MHD modes ($n\lesssim20$) with micro-turbulence [$n\sim \mathcal{O}(100)$], we apply magnetic field-aligned Fourier filters to separate modes with low and high toroidal mode numbers, using the toroidally spectral solver for the low-$n$ part of the spectrum, and the large aspect ratio solver for the high-n part.
This approach limits the number of unknowns per field equation and poloidal plane to a maximum of three times the number mesh vertices per plane, significantly less compared to using the spectral solver alone.

In Sec. \ref{sec:error_analysis}, we carried out a quantitative error analysis comparing the new spectral field solver to the conventional large aspect ratio solver.
Our analysis confirms that the large aspect ratio solver is adequate for micro-turbulence in tokamak edge plasma ($k_\parallel \ll k_\perp$) even in tight aspect ratio scenarios like NSTX.
Reduced-MHD modes, however, occur at low toroidal and poloidal mode numbers and exhibit (unacceptably) large errors reinforcing our assessment that a more accurate field solver was necessary at least for the low-$n$ part of the spectrum.

We verified the new field solver for several basic physics effects in Sec. \ref{sec:verification}.
We calculated the frequency and damping rate of shear-Alfv\'{e}n waves and successfully matched the results of the analytic dispersion relation and another version of XGC using  the large aspect ratio approximation in its field solver.
We observed good agreement between XGC, the gyrokinetic code GTS, and the MHD code NIMROD for the $(n,m)=(1,1)$ collisionless kink mode, and between XGC and ORB5 for the $(n,m)=(1,2)$ collisionless tearing mode.
We also verified the new field solver for nonlinear electromagnetic turbulence in a benchmark against XGC using the large aspect ratio approximation in the field equations.

With the improved field solver introduced here, XGC is now ready for studying reduced-MHD modes and their interaction with micro-turbulence.
The new capability is essential for investigating problems such as how micro-turbulence affects the stability of ELMs, or reacts to magnetic islands due to the presence of tearing modes, and the gyrokinetic plasma response to RMP fields.
Future work will include a benchmark of peeling-ballooning mode physics against an MHD code such as M3D-C1 \cite{Jardin2004,Breslau2009}.

\roles{
R. Hager: Writing -- original draft, writing -- review and editing, methodology, software, investigation, visualization, validation, formal analysis.
C. S. Chang: Supervision, project administration.
T. Gade: Data curation, investigation, writing -- review and editing.
E. Held: Data curation, validation, writing -- review and editing.
S. Ku: Software, writing -- review and editing.
A. Mishchenko: Data curation, investigation, writing -- review and editing.
A. Scheinberg: Software, data curation.
B. Sturdevant: Formal analysis, methodology, data curation, writing -- review and editing.
}

\ack{The authors would like to thank Edward Startsev, Yang Chen and Scott Parker for fruitful discussions.}

\funding{
This work was supported by the U.S. Department of Energy under contract number DE-AC02-09CH11466 (PPPL) [including subcontracts P-240001662 to Utah State University (E. Held) and P-240000447 to Jubilee Development (A. Scheinberg)] through the Scientific Discovery through Advanced Computing (SciDAC) program.
The United States Government retains a non-exclusive, paid-up, irrevocable, world-wide license to publish or reproduce the published form of this manuscript, or allow others to do so, for United States Government purposes.
Thomas Gade received funding through the U.S. Department of Energy, Office of Science, Office of Advanced Scientific Computing Research, Department of Energy Computational Science Graduate Fellowship under Award Number DE-SC0023112.
This research used resources of the National Energy Research Scientific Computing Center (NERSC), a Department of Energy Office of Science User Facility using NERSC award FES-ERCAP0027958.
}

\data{Data used in preparing this article will be available in the PPPL Theory Department Data Repository, see Ref. \onlinecite{xgc_data}.}
%



\bibliographystyle{iopart-num}
\bibliography{references}

\end{document}